\documentclass[aps,prb,twocolumn,superscriptaddress,showpacs,floatfix,amsmath,amssymb]{revtex4-2}
\usepackage{graphicx}
\usepackage{dcolumn}
\usepackage{bm}
\usepackage{amsmath} 
\usepackage[below]{placeins}
\usepackage{setspace} 
\usepackage{multirow}
\usepackage{xcolor}

\newcommand{\ruby}{{(ND$_{4}$)$_{2}$FeCl$_{5}\cdot$D$_{2}$O}}
\newcommand{\rubyH}{{(NH$_{4}$)$_{2}$FeCl$_{5}\cdot$H$_{2}$O}}

\newcommand{\red}[1]{{\color{black} #1}}

\begin{document}
\begin{titlepage}

\pagebreak[4]
\end{titlepage}

\title{Magnetic Excitations of the Hybrid Multiferroic {{\ruby}}}
\thanks{This manuscript has been authored by UT-Battelle, LLC under Contract No. DE-AC05-00OR22725 with the U.S. Department of Energy.  The United States Government retains and the publisher, by accepting the article for publication, acknowledges that the United States Government retains a non-exclusive, paid-up, irrevocable, world-wide license to publish or reproduce the published form of this manuscript, or allow others to do so, for United States Government purposes.  The Department of Energy will provide public access to these results of federally sponsored research in accordance with the DOE Public Access Plan (http://energy.gov/downloads/doe-public-access-plan).\\}

\author{Xiaojian Bai}
\email{baix@ornl.gov} \affiliation{Neutron Scattering Division,
Oak Ridge National Laboratory, Oak Ridge, Tennessee 37831, USA}

\author{Randy S. Fishman}
\affiliation{Materials Science and Technology Division, Oak Ridge
National Laboratory, Oak Ridge, Tennessee 37831, USA}

\author{Gabriele  Sala}
\affiliation{Neutron Scattering Division, Oak Ridge National
Laboratory, Oak Ridge, Tennessee 37831, USA}

\author{Daniel M. Pajerowski}
\affiliation{Neutron Scattering Division, Oak Ridge National
Laboratory, Oak Ridge, Tennessee 37831, USA}

\author{V. Ovidiu Garlea}
\affiliation{Neutron Scattering Division, Oak Ridge National
Laboratory, Oak Ridge, Tennessee 37831, USA}

\author{Tao Hong}
\affiliation{Neutron Scattering Division, Oak Ridge National
Laboratory, Oak Ridge, Tennessee 37831, USA}

\author{Minseong Lee}
\affiliation{National High Magnetic Field Laboratory, Los Alamos National Laboratory, Los Alamos, New Mexico 87545, USA}

\author{Jaime A. Fernandez-Baca}
\affiliation{Neutron Scattering Division, Oak Ridge National Laboratory, Oak Ridge, Tennessee 37831,
USA}

\author{Huibo Cao}
\affiliation{Neutron Scattering Division, Oak Ridge National
Laboratory, Oak Ridge, Tennessee 37831, USA}

\author{Wei Tian}
\email{wt6@ornl.gov} \affiliation{Neutron Scattering Division,
Oak Ridge National Laboratory, Oak Ridge, Tennessee 37831, USA}
\date{\today}

\begin{abstract}

We report a comprehensive inelastic neutron scattering study of the hybrid molecule-based multiferroic compound {{\ruby}} in the zero-field incommensurate cycloidal phase and the high-field quasi-collinear phase. The spontaneous electric polarization changes its direction concurrently with the field-induced magnetic transition, from mostly aligned with the crystallographic $a$-axis to the $c$-axis. To account for such change of polarization direction, the underlying multiferroic mechanism was proposed to switch from \red{the spin-current model induced via} the inverse Dzyalloshinskii--Moriya interaction to the $p$-$d$ hybridization model \cite{rodriguez2017magnetic}. We perform a detailed analysis of the inelastic neutron data of {\ruby} using linear spin-wave theory \red{to quantify magnetic interaction strengths and investigate possible impact of different multiferroic mechanisms on the magnetic couplings}. Our result reveals that the spin dynamics of both multiferroic phases can be well-described by a Heisenberg Hamiltonian with an easy-plane anisotropy. 
We \red{do not} find notable differences between the optimal model parameters of the two phases. The hierarchy of exchange couplings and the balance among frustrated interactions \red{remain the same} between two phases, suggesting that magnetic interactions in {\ruby} are much more robust than the electric polarization in response to delicate reorganizations of the electronic degrees of freedom in an applied magnetic field.

\end{abstract}


\maketitle
\section{Introduction}
Crystalline condensed matter systems are assemblies of lattice, charge and spin degrees of freedoms with a hierarchy of complex interactions. In thermal equilibrium, materials may display various linear responses to small external stimuli, such as strain to stress, polarization to electric fields and magnetization to magnetic fields. A cross coupling of polarization (magnetization) with a magnetic (electric) field is generally referred as the magnetoelectric effect \citep{fiebig2005revival}. Single-phase magnetoelectric materials that host both ferroelectric order and (anti)ferromagnetic order are called multiferroics \citep{schmid1994multi}. Understanding microscopic instabilities that give rise to these coupled phenomena is of fundamental interest.

Ferroelectric order requires breaking the spatial inversion symmetry. This can be achieved by a polar distortion driven by a structural instability, such as in BiFeO$_3$ where the major source of polarization is produced by  Bi$^{3+}$ displacements \citep{kubel1990structure}. The long-range magnetic ordering in this material occurs on a different site \citep{sosnowska1982spiral}, Fe$^{3+}$, and at a temperature $\sim 460$~K lower than the ferroelectric ordering $T_{\text{FE}}\approx 1103$~K. Although two types of order coexist in BiFeO$_3$, they are of completely different origins and separated in energy scales. As a result, the coupling between them is quite weak. This is the typical case for so-called type-I (or proper) multiferroics. In comparison, the type-II (or improper) multiferroics \citep{Cheong-2007, Khomskii-2009}, such as TbMnO$_3$, are known for their strong magnetoelectric coupling effect which is driven by correlated electronic degrees of freedom. The spatial inversion symmetry in TbMnO$_3$ is broken by the formation of a cycloidal spin structure and the electric polarization is created by the lattice response to magnetic frustration through the inverse Dzyaloshinskii--Moriya (DM) mechanism \citep{walker2011femtoscale,katsura2005spin,Dagotto-2006}. Naturally, in this case, ferroelectric ordering is strongly coupled with magnetic ordering. The direction of the electric polarization is completely switchable by rotating the spiral plane in an applied magnetic field \citep{TbMnO3-nature-2003}. Meanwhile, chiral magnetic domains can also be selected by poling electric fields \citep{yamasaki2007electric}. However, the absolute magnitude of the polarization in TbMnO$_3$ is rather small \citep{TbMnO3-nature-2003}, only $\sim 0.1\%$ of BiFeO$_3$, and the transition takes place at a much lower temperature, $T_{\text{FE}}\approx 28$~K.

The intimate coupling between ferroelectricity and magnetism in multiferroics provides a new knob to tune materials across quantum phase transition \citep{narayan2019multiferroic}, realizing novel emergent phases. The progress of multiferroic research is heavily driven by discoveries of new materials. Most known multiferroic compounds are transition metal oxides, such as TbMnO$_3$ \citep{TbMnO3-nature-2003, yamasaki2007electric,Dagotto-2006, Xiang-2008, Malashevich-2008, TbMnO3-Kenzelmann-2005,Lovessey-2013,Solovyev-2011}, MnWO$_4$ \citep{MnWO4-2006}, Ni$_3$V$_2$O$_8$ \citep{Ni3V2O8-2005}, CuO \citep{CuO-2012} and LiCuVO$_4$ \citep{Xiang-2007}, to name some that have spiral spin order \citep{wang2009multiferroicity}. The discovery of multiferoic behavior in crystalline molecule-based magnets and metal-organic framework materials (MOFs) has significantly expanded our horizon \citep{Omar-2003, Saman-2012, Xu-2011, Samantaray-2011}. Structurally, such hybrid magnets consist of ``hard'' polyhedron building blocks bridged by ``soft'' organic linkers. This unique combination makes their magnetic properties highly tailorable. One can manipulate interaction pathways, spin anisotropy and effective dimensionality via ligand engineering. Interesting new phases can be readily stabilized with modest fields and pressures. Therefore, hybrid molecule-based multiferroics are an excellent platform to explore new physics, as well as to develop next-generation multi-functional devices.

The linear magnetoelectric response was recently discovered in the molecule-based compounds $A_{2}$[FeCl$_{5}\cdot$H$_{2}$O], $A=$ K, Rb or Cs \citep{Ackermann-2014}. The crystal structure of this family of compounds consists of distorted [FeCl$_{5}\cdot$H$_{2}$O]$^{2-}$ octahedra linked by a network of hydrogen bonds \citep{Carlin-1977}. A single magnetic ordering transition was observed in all three compounds at temperatures ranging from $14$ to $4.5$~K. Electric polarization can be induced in the ordered phases by applying an external magnetic field \citep{Ackermann-2014}. Interestingly, these materials transform into a multiferroic when the alkali metals are replaced by ammonium cations [NH$_4$]$^{-}$ \citep{Ackermann-2013}. In this case, electric polarization emerges \textit{spontaneously} in zero field at the onset of the low-temperature cycloidal magnetic ordering -- a key difference with linear magnetoelectric materials. Comprehensive thermodynamic measurements reveal rich multiferroic phase diagrams in an applied magnetic field along the crystallographic $a$- and $c$-axis, while the $b$-axis is the magnetic hard axis \citep{Ackermann-2013, clune2019magnetic}. The magnetic structures in different phases were solved from single crystal neutron diffraction using deuterated samples \citep{Jose-2015, rodriguez2017magnetic, Tian-PRB, bruning2020magnetoelectric}. 

\begin{figure} [tp!]
\centering\includegraphics[width=1\columnwidth]{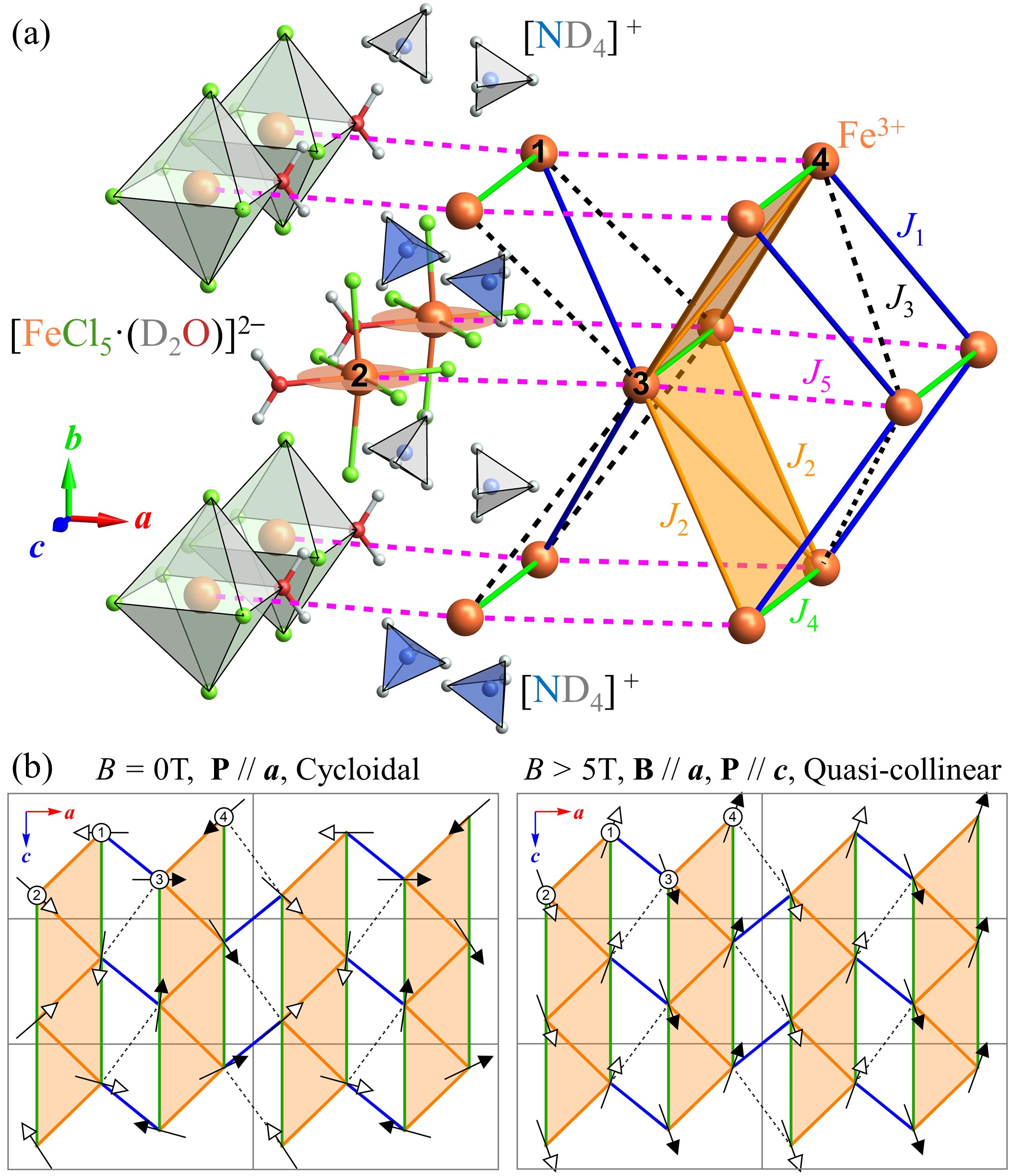}
\caption{\label{fig1}  Low temperature crystal and magnetic structures determined by neutron diffraction study using deuterated {\ruby}. (a) Partial crystal structure and magnetic exchange pathways of {\ruby} below the structural transition at $T_{\text{S}}\approx 79$~K. The transition is associated with an orientational ordering of [ND$_4$]$^{-}$ tetrahedra. Two symmetry-inequivalent groups of [ND$_4$]$^{-}$ are rendered in gray and blue, respectively. Three dominant exchange interactions $(J_1,J_2,J_4)$ are plotted as solid lines, and two sub-leading ones $(J_3,J_5)$ are plotted as dashed lines. Antiferromagnetic $J_2$ and $J_4$ bonds form a buckled frustrated triangular-lattice planes (orange). Octahedrally coordinated local environment of Fe$^{3+}$ atom produces a weak easy-plane magnetic anisotropy, which coincides with the ac-plane and is illustrated by orange disks on one of the four Fe$^{3+}$ sublattices. (b) The low temperature magnetic structures of {\ruby} in zero field (cycloidal, ${\bf k} = (0,0,0.23)$) and high fields (quasi-collinear, ${\bf k} = (0,0,0)$), projected onto the ac-plane. The  transverse cycloids of spins are stabilized by competing $J_2$ (orange) and $J_4$ (green) exchanges on the buckled triangular-lattice plane. The empty and filled arrows represent spins on two parallel cycloids. Magnetic moments on sublattice 1 and 3 in the same unit cells are strictly antiparallel (solid grey lines are unit cell edges), so tthe moments on sublattices 2 and 4 across neighboring unit cells are favored by the dominant antiferromagnetic $J_1$ (blue) couplings. The subleading interaction $J_5$ is omitted for clarity. A spin-flop transition leads to the canted antiferromagnetic structure when an external field ($\approx 5$~T) is applied along $\textit{\textbf{a}}$ at $T=2$~K. The electric polarization switches from mostly parallel to $\textit{\textbf{a}}$ to parallel to $\textit{\textbf{c}}$ after the transition.}
\end{figure}

{{\rubyH}} undergoes three transitions \textit{en route} to multiferroicity in zero field. First, a structural transition occurs at $T_\text{S}\approx 79$~K. The crystal symmetry is lowered from orthorhombic $Pnma$ to monoclinic $P112_1/a$ with detectable $\sim 0.1^\circ$ change in one of the cell angles. This change is associated with an orientational ordering of [NH$_4$]$^{+}$ groups \citep{bruning2020magnetoelectric}, Fig.~\ref{fig1}(a). Next, a long-range magnetic ordering takes place at $T_{N}\approx 7.3$~K, stabilizing a collinear sinusoidal structure with magnetic moments along the $a$-axis \citep{rodriguez2018switching}. At $T_{\text{FE}}\sim 6.8$~K, the magnetic structure changes to an incommensurate cycloid in the ac-plane with propagation vector ${\bf k} = (0,0,0.23)$, Fig.~\ref{fig1}(b). At the same time, a spontaneous electric polarization emerges along ${\textbf{\textit{a}}}$. The magnitude of this polarization is quite small at $T=3$~K, only $\sim 1\%$ of the polarization in TbMnO$_3$. The direction of the polarization is consistent with the prediction of the inverse DM mechanism, ${\bf P}\sim {\bf r}_{ij}\times({\bf S}_{i}\times {\bf S}_j)$, where ${\bf S}_{i}\times {\bf S}_j//{\textbf{\textit{b}}}$ and $ {\bf r}_{ij}//{\textbf{\textit{c}}}$, therefore ${\bf P}//{\textbf{\textit{a}}}$.

A spin-flop transition occurs in the multiferroic phase of {\rubyH} at $B\approx 5$~T when the field is applied in the $a$-axis, resulting in a ${\bf k}={\bf 0}$ quasi-collinear magnetic structure, Fig.~\ref{fig1}(b). Similar transition happens at a lower field of $B\approx 3.5$~T when field is applied along the $c$-axis \citep{Ackermann-2013, clune2019magnetic}. In both cases, the electric polarization rotates to the $c$-axis and increases linearly up to the highest field of $B=14$~T in the measurement. The spin-current model can no longer explain the observed electric polarization in these spin-flop phases. Surprisingly, a spin-dependent $p$-$d$ hybridization model produce a non-vanishing polarization along the $c$-axis for fields applied in either the $a$- or $c$-axis, based on the local environment of Fe$^{3+}$ and the canted spin structure \citep{rodriguez2017magnetic}. 

It is rather unusual to have two unrelated mechanisms active in a single material. If the multiferroic mechanism switches, does it leave a fingerprint in the microscopic magnetic interactions? In a broader context, we often think of the exchange interactions as ``constants", but in reality the lattice may be distorted in order to relieve magnetic frustration, which in turn affects magnetic interactions through the complex interplay of various degrees of freedom in the material. Different multiferroic phases of {\rubyH} are readily accessible at modest fields and temperatures, providing us an excellent example to explore such influence. Moreover, a pressure-induced multiferroic reentrant behavior was recently discovered in deuterated {\ruby} \citep{wu2021reentrance}. Our work on quantifying the magnetic interactions at zero pressure provides a basis for understanding such exotic behavior.

\section{EXPERIMENTAL DETAILS}

To determine the microscopic exchanges in the different multiferroic phases, we map out the entire energy and momentum dependence of excitation spectra in zero field and $B=6$~T measuring a large deuterated single crystal sample of {\ruby} using neutrons. Deuterated sample is needed for inelastic neutron scattering measurements to reduce the incoherent scattering background from hydrogen \citep{Tian-PRB}. A large single crystal of $0.6$\,gram was mounted on an aluminum plate and aligned in the $(0,k,l)$ scattering plane using the CG-1B alignment station at the High Flux Isotope Reactor (HFIR), Oak Ridge National Laboratory (ORNL), USA.

Two separate non-polarized experiments were performed using the same mount in zero field and applied fields on the Cold Neutron Chopper Spectrometer (CNCS) at the Spallation Neutron Source (SNS), ORNL, USA \citep{ehlers2011new}.
An incident energy of $E_i=3.32$~meV in the high flux mode 
was used in the CNCS experiments, yielding an elastic energy resolution (FWHM) of $\approx 0.12$~meV.  A liquid-helium cryostat was used to cool the sample down to the base temperature of $T\approx 2$~K in both zero-field and field experiment. An 8~T vertical cryomagnet was used in the latter experiment to supply a field up to $B=6$~T along the $a$-axis of the sample. An instrumental background dataset was collected at $T=20$~K with cryomagnet and subtracted out in data processing, Fig.~S3. No background subtraction was applied to the zero-field data. The raw event data were symmetrized according to the $mmm$ point group of the high-temperature space group $Pnma$ and converted into a histogram format in Mantid \citep{arnold2014mantid}. During this process, a detector efficiency correction using a Vanadium standard was applied. \red{The time of flight in the zero-field experiment was corrected by a time offset to account for the shift of elastic line, Fig.~S2.}   The histogram data were imported into Horace \citep{ewings2016horace} and modeled using the linear spin-wave theory (LSWT) \citep{Randy-book} and SpinW \citep{toth2015linear}.

\begin{figure*}[t]
\centering\includegraphics[width=0.85\textwidth]{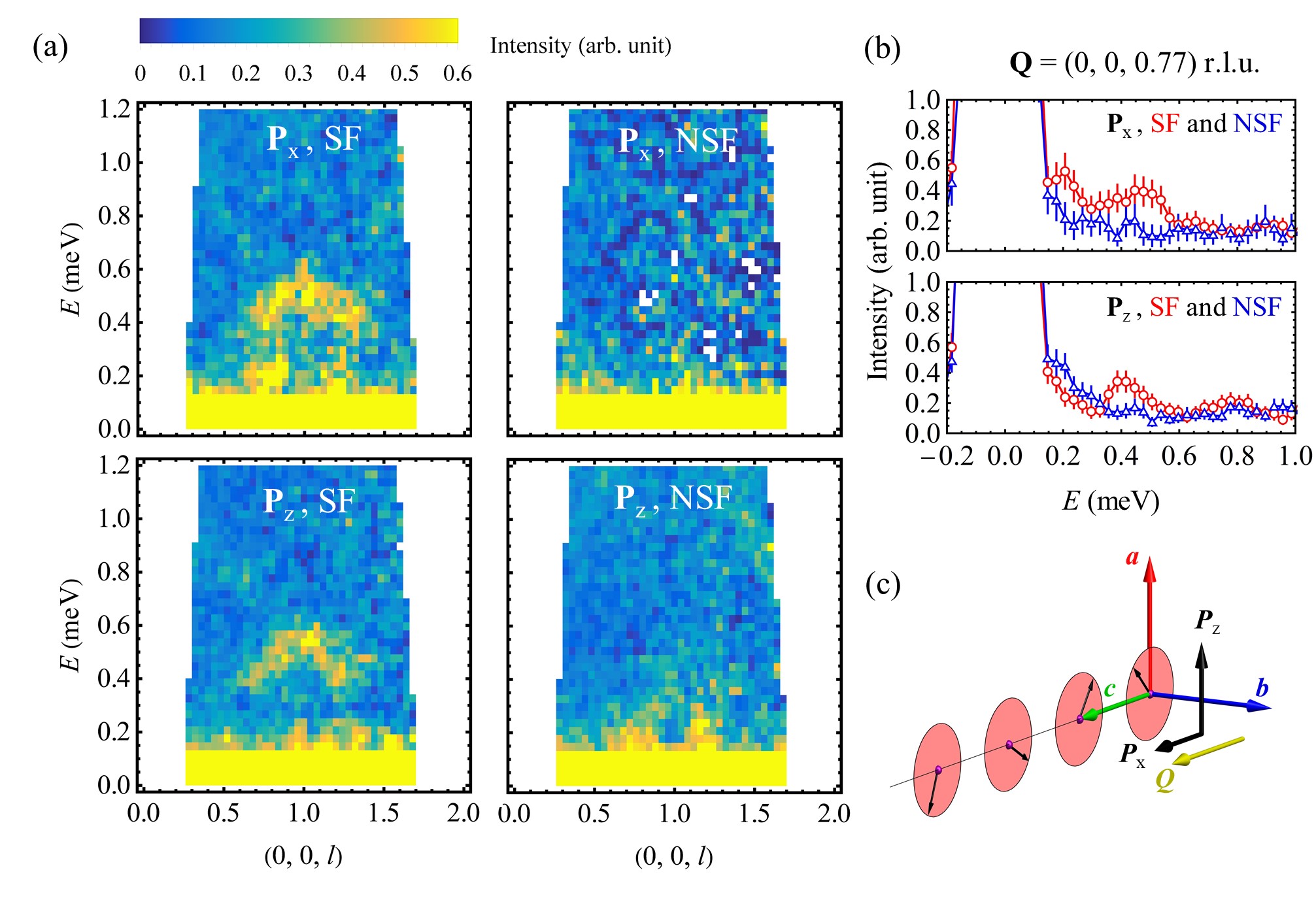}
\caption{\label{polarized} Polarized neutron scattering data of {\ruby}  measured at $T=2$\,K with incident neutron energy $E_\text{i}=3.8$~meV on the HYSPEC instrument. (a) Energy-momentum slices in the spin-flip (SF) and non-spin-flip (NSF) channel with neutron polarized along the momentum transfer (${\bf P}_x$ $\parallel {\bf c}$) and perpendicular to the scattering plane (${\bf P}_z$ $\parallel {\bf a}$). (b) Constant-$\bf Q$ cuts at the incommensurate ordering wave-vector. (c) Schematic plot of the scattering geometry in the polarized neutron experiment, showing the cycloidal plane (pink disk), polarization of incoming neutrons (black arrows) and scattering wave-vector (yellow arrow).}
\end{figure*}

Complementary polarized neutron scattering measurements were carried out on the Hybrid Spectrometer (HYSPEC) at SNS, ORNL, USA \citep{stone2014comparison,winn2015recent}. An incident energy of $E_i=3.8$~meV and Fermi Chopper frequency $f = 180$~Hz were used in the experiment, yielding an elastic energy resolution (FWHM) of $\approx 0.12$~meV. The incident neutron beam was polarized using a vertically focusing Heusler monochromator, and the outgoing beam was analyzed using a radially collimating supermirror array. 
The flipping ratio was $\sim 11$ based on measurements of the $(0,2,0)$ nuclear Bragg peak. Additional neutron scattering data were collected on the Cold Neutron Triple-Axis Spectrometer (CTAX) at HFIR, ORNL, USA [Supplementary Section S1].

\section{MAGNETIC SPECTRUM AND SPIN-WAVE MODELING}

Before diving into the detailed modeling of the magnetic interactions in {\ruby}, we first use the polarized neutron technique to examine the excitation spectrum in the zero-field cycloidal phase and characterize the nature of different magnetic modes. Our main observations are summarized in Fig.~\ref{polarized}. \red{The low energy window ($<1.2$~meV) is shown to highlight the excitation spectrum in different polarization channels. There is no magnetic signal at higher energies for this cut, see Fig.~3(a).} The usual selection rule requires that only magnetic fluctuations perpendicular to the scattering wave-vector {$\bf Q$} can be detected by neutrons. In addition to this rule, the spin-flip (SF) channel in polarized neutron scattering experiments is sensitive only to magnetization components perpendicular to neutron polarization, while the non-spin-flip (NSF) scattering is sensitive only to components parallel to neutron polarization. In the ${\bf P}_x$ mode where neutron is polarized along the scattering wave-vector ${\bf Q}$, scattering intensities only appear in the SF channel confirming the magnetic nature of the excitation spectra. In the ${\bf P}_z$ mode where neutron polarization is perpendicular to ${\bf Q}$ and the scattering plane, the scattering signals clearly split into two parts, the optical branches in the SF channel and the acoustic branches in the NSF channel. The former corresponds to fluctuations out of the cycloidal plane and along the $b$-axis. These modes are gapped due to additional energy cost in overcoming the easy-plane anisotropy. The latter corresponds to fluctuations along the $a$-axis. They are gappless modes coming from collective rotation of spins within the easy-plane, also known as phason modes.

\begin{figure*}[t!]
\centering\includegraphics[width=1\textwidth]{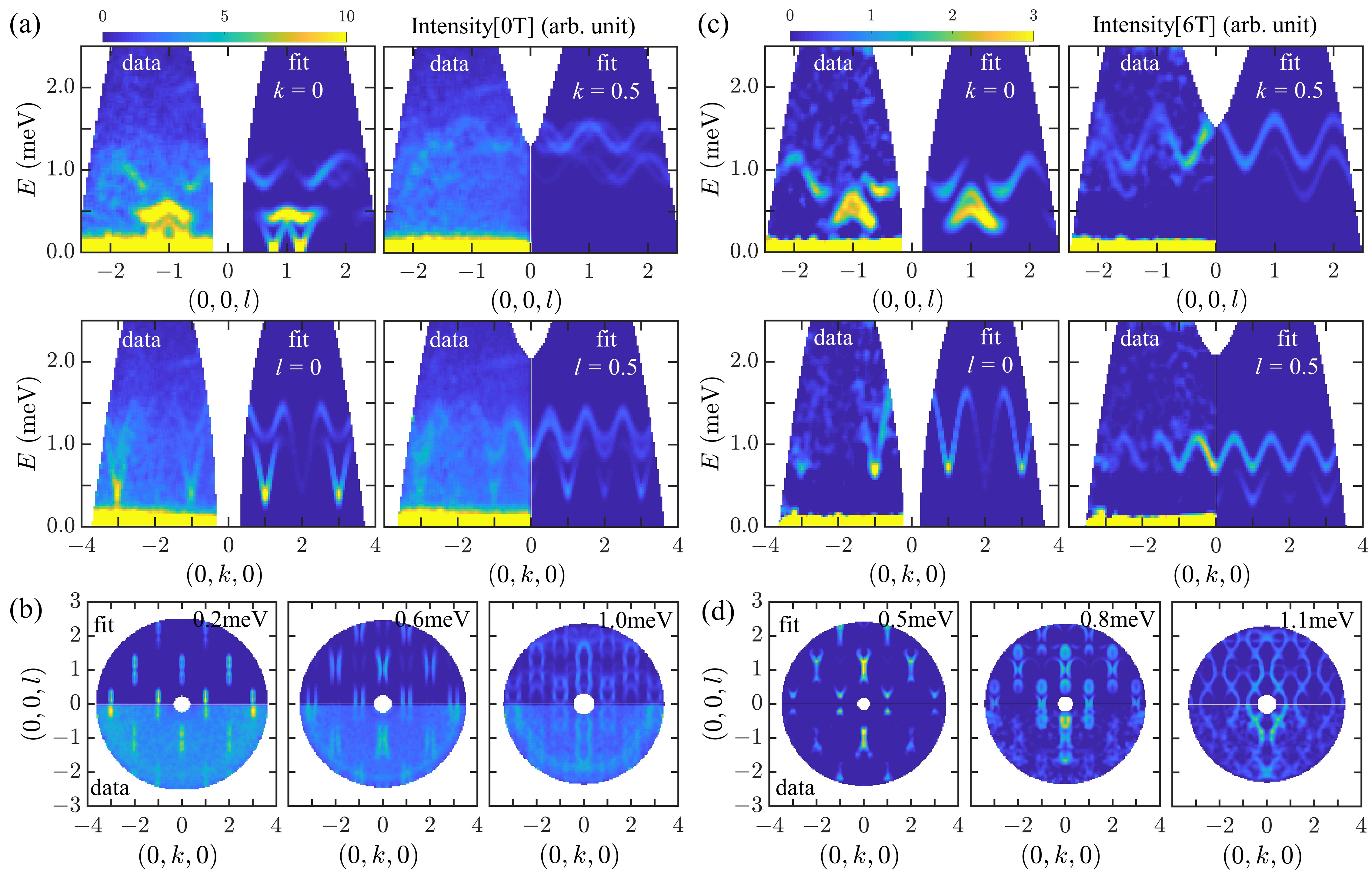}
\caption{\label{fig:inelastic} Magnetic excitations of {\ruby}  measured at $T=2$\,K in $B=0$~T and $6$~T applied field  with incident neutron energy $E_\text{i}=3.32$~meV on the CNCS instrument, and comparison with respective best fits using linear spin-wave theory (LSWT). (a) Representative energy-momentum slices in the $(0,k,l)$-plane of the zero-field data and its best fitted model. An integration over $|\Delta l|\le 0.05$~r.l.u. ($|\Delta k|\le 0.05$~r.l.u.) is performed for constant-$k$ (-$l$) cuts. (b) Momentum-dependence of excitation spectra of the zero-field data and the model integrated over $\pm0.05$~meV at selected energies. (c)(d) Similar plots for the $B=6$~T data and its best fitted model. \red{Note that the discrepancy between the data and modeling for the $l=0.5$~r.l.u. data at $6$~T can be attributed to background over-subtraction as illustrated in Fig.~S3.}  An integration over $|h|\le 0.05$~r.l.u. is applied throughout all panels. The fitted \red{linear} background functions are not plotted for clarity.}
\end{figure*}
\setlength{\tabcolsep}{6pt}
\renewcommand{\arraystretch}{1.2}
\begin{table*}[]
\centering
\begin{tabular}{c  c  c  c  c  c  c  c  c  c  c}
\hline\hline
$B$ (T) & $J_1$ (meV) & $J_2$       & $J_3$       & $J_4$       & $J_5$      & $D$        & $\chi^2_\text{red}$ &  $k_c $ & $\phi$ ($^\circ$) & $\theta$ ($^\circ$) \\ \hline
0.0       & 0.177(10)   & 0.064(5)  & 0.029(7)  & 0.056$^\ddag$  & 0.035$^\ddag$ & 0.015(7) & 22.96                        & 0.23$^\dagger$       & 41.50$^\dagger$         &        12.02$^*$        \\ \hline
6.0       & 0.181(13)   & 0.051(3) & 0.033(4) & 0.055(3) & 0.039(7) & 0.014(6) &    1.31                &  0.24$^*$  &  40.25$^*$       & 12.66           \\ \hline\hline
\end{tabular}
\caption{Values of the best fitting parameters for {\ruby} in two different multiferroic phases. $^{\dagger}$ parameters, the incommensurate ordering wave-vector $k_c$ and the phase difference $\phi$ between spin on sublattice 1 and 4, are fixed in fitting of the zero-field data. $\ddag$ parameters, $J_4$ and $J_5$, are calculated on-the-fly using Eq.~\eqref{eq:J4} and \eqref{eq:J5}. $^*$ values are computed using respectively optimal fitting parameters. The large reduced $\chi^2_\text{red}$ for the zero-field data is due to incoherent backgrounds that can not be fully accounted for \red{using a simple linear background function}. The parameter $\theta$ is the canting angle of spin on sublattice 2 from the crystallographic $c$-axis. \red{The error is estimated by the change of a parameter that produces $5\%$ increase of the $\chi^2_\text{red}$ with all the rest of parameters fixed at optimal values. The two models are consistent within the estimated errors.}  }
\label{tab1}
\end{table*}

Magnetic couplings in {\ruby} are quite complex. The exchange pathways up to the fifth nearest neighbors are shown in Fig.~\ref{fig1}. To make progress in quantifying magnetic interactions in {\ruby}, we keep complexity of the model to a minimum and introduce a Heisenberg Hamiltonian in a magnetic field with an easy-plane anisotropy,
\begin{equation}
{\mathcal H} = \sum_{i<j} J_{ij}{\mathbf S}_{i}\cdot {\mathbf S}_{j} + D \sum_{i} \left(S_{i}^{b}\right)^2+gB\sum_i S_{i}^{a},
\label{eq:spinhamiltonian}
\end{equation}
where ${\bf S}_i$ is the spin operator of Fe$^{3+}$ ion on site $i$ with length $S=5/2$, $D$ is the single-ion anisotropy and $g =2$ is the electron g-factor.
We perform independent pixel-to-pixel fittings to a volume of 4-dimensional $(h, k,l,E)$ data \red{with $0\le h\le 0.5$~r.l.u., $0\le k \le 2$~r.l.u., $0\le l \le 2$~r.l.u., $0.25\le E\le 2$~meV}, collected in the zero-field cycloidal phase, and a volume of 3-dimensional $(k,l,E)$ data \red{with $0\le k\le 2$~r.l.u., $0\le l \le 2$~r.l.u., $0.25\le E\le 2$~meV}, collected in the canted antiferromagnetic phase of $B=6$~T. Due to the narrow opening of the vertical magnet, the momentum transfer along $h$ is limited and integrated over $|h|<0.05$~r.l.u. in analysis of the field data. A simple linear background, \red{$I_\text{bk}({\bf Q},E) = a_\text{bk}E+b_\text{bk}$}, is included in fitting each dataset. The calculated spectra are convoluted with instrumental resolution (FWHM~$\approx0.12$~meV). 

\begin{widetext}
\begin{align}
&\frac{E}{NS^2}=-8J_2\cos (2\pi k_c/2) +4J_4\cos(2\pi k_c)
- 4J_1 \cos(2\pi k_c/2+\phi) 
-4J_3 \cos(2\pi k_c/2-\phi ) + 4J_5 \cos (\phi )\label{eq:E}\\
&J_4 = \dfrac{1}{2\sin(2\pi k_c)}\left(2J_2\sin\left(2\pi k_c/2\right)+J_1\sin\left(2\pi k_c/2+\phi\right)+J_3\sin\left(2\pi k_c/2-\phi\right)\right)\label{eq:J4}\\
&J_5 = (J_1-J_3)\sin\left(2\pi k_c/2\right)/\tan(\phi)+(J_1+J_3)\cos\left(2\pi k_c/2\right)\label{eq:J5}
\end{align}
\end{widetext}

\begin{figure}[t]
\centering\includegraphics[width=1\columnwidth]{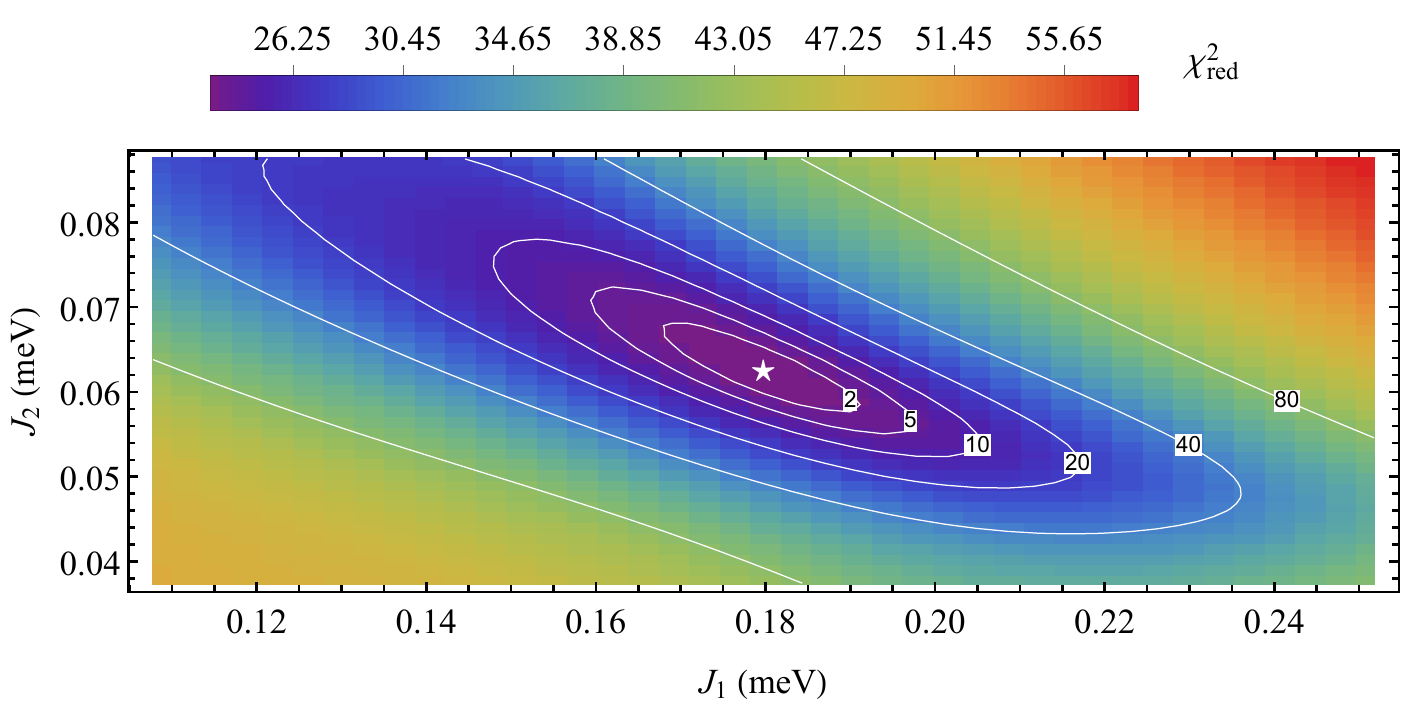}
\caption{\label{fig:scan_chisq} \red{A $\chi^2_\text{red}$-map in the parameter space of $J_1$ and $J_2$ with all the other parameters fixed at optimal values. The star indicates the optimal fit. The numbers on the white contours label percentage increases from the minimal value of $\chi^2_\text{red}$.}}
\end{figure}

In zero field, we obtain two analytic constraints of exchange parameters, Eq.\eqref{eq:J4} and \eqref{eq:J5}, by minimizing the energy, Eq.\eqref{eq:E}, with respect to the incommensurate wave vector $k_c$ and the phase difference $\phi$ between sublattice 1 and 4. 
The exchange parameter $J_4$ and $J_5$ are calculated using these constraints in every iteration of the fitting procedure, so that the correct magnetic ground state is enforced. The source of frustration comes from antiferromagnetic $J_2$ and $J_4$ on the buckled triangular-lattice plane. Setting all the other exchanges to zero, we recover the well-known result for stabilizing a spiral structure with $2\pi k_c = 2\arccos(J_2/(2J_4))$ in the Heisenberg model on an anisotropic triangular lattice \citep{lyons1960method}. The relative phase difference $\phi$ of cycloids on different sublattices in {\ruby} is tuned by the exchange parameters $J_1, J_3$ and $J_5$. \red{Fittings are performed for $k_c=0.23$~r.l.u. and various values of the phase difference $\phi$ ranging from $30^\circ$ to $48^\circ$, see Fig.~S4 for details. The result shows that our inelastic fitting is sensitive to the phase difference in the cycloidal ground state and yields an optimal value $\sim 39^\circ$, which reasonably agrees with the experimental value, $41.5^\circ$, determined from neutron diffraction \citep{Jose-2015}. The fitted exchange model parameters are reported in Tab.\ref{tab1} for the case of $\phi=41.5^\circ$. To avoid any bias in determining the model parameters in high-field phase, no constraints are applied in fitting the $B=6$~T data. The comparison between experimental data and corresponding fits is presented in Fig.~\ref{fig:inelastic} for selected cuts. See Supplementary Section~S4 for more details. The agreement is fairly good for both phases across the whole reciprocal space detected in the experiments. 
From the parameter correlation matrix given in Tab.~S1 and~S2, we find that $J_1$ and $J_2$ are negatively correlated to a high degree, $0.881$, in both models. To investigate such correlation, we map out $\chi^2_\text{red}$ for the zero-field model in the parameter space of $J_1$ and $J_2$ with all the other parameters fixed at optimal values. The result in Fig.~\ref{fig:scan_chisq} reveals an elongated region with small $\chi^2_\text{red}$, which is the source of large correlations between the two parameters. However, it is important to note that we do not have a line of degeneracy in the parameter space. Moreover, the elongated contours smoothly converge toward the point of optimal fit indicated by the star. To quantify the parameter uncertainty in the local region of parameter space, we report the errors in Tab.~\ref{tab1} as the change of a parameter that produces $5\%$ increase of the $\chi^2_\text{red}$ with all the rest of parameters fixed at optimal values. 
Within this error estimation, we can confidently conclude that our model is a unique local minimum. 
To further examine if our model is the global minimum, we performed 100 fittings starting with randomized initial values within $\pm 50\%$ of the optimal fits. Majority of fittings (79 cases) converges and yields the same set of optimal parameters with $\le 5\%$ of standard deviations. The rest of fits either find local minimums with higher reduced $\chi^2_\text{red}$ (6 cases) or generates imaginary modes in the LSWT calculations (15 cases) due to incompatibility with the experimental cycloidal ground state. Global minimum analysis is also performed for the high-field model and yield similar results.}
 
\section{DISCUSSION AND CONCLUSION}
\red{Two important conclusions can be drawn from our detailed LSWT modeling. First, the zero-field and high-field inelastic data can be satisfactorily described by essentially the same model within the estimated parameter errors. 
We further cross-check the incommensurate ordering wave-vector $k_c$ and the phase difference $\phi$ of the $0$~T cycloidal phase using the $6$~T model parameters and do not find significant difference from the experimental values (fixed in fitting the zero-field data). Vice versa, we compute the canting angle $\theta$ at $6$~T using the $0$~T model and obtain almost identical result as the $6$~T model. We also find the saturation magnetic field to be $\sim 27.5$~T for both models, close to the experimental value of $B_{\text{sat}}\approx 30$~T \citep{clune2019magnetic}.
In light of the evidence that the underlying mutiferroic mechanism switches from the spin-current model to the $p$-$d$ hybridization model under magnetic field \cite{rodriguez2017magnetic}, it is quite surprising that such change leaves very little dynamical fingerprint in the magnetic excitation spectra.
Second, the fitted values of $J_2$ and $J_4$ are very close, indeed giving rise to expected magnetic frustration. This is in sharp contrast with models of K$_2$FeCl$_5\cdot$H$_2$O extracted from inelastic neutron data \citep{Campo-INS-2008}, where $J_4$ is significantly smaller than $J_2$.  Therefore, a collinear structure instead of cycloid is stabilized at zero field in this compound \citep{Gabas-1995}. The presence of ammonium groups is closely related to the enhancement of $J_4$ and the magnetic frustration in the system. A recent neutron diffraction study under pressure suggests that a moderate external pressure may alter the conformation of ammonium groups and produce novel multiferroic reentrant phenomenon in {\ruby} \citep{wu2021reentrance}. }

\red{Finally, we discuss several subtle aspects that are not taken into account in our current modeling. At room temperature, the only symmetry of the Fe$^{3+}$ site is a mirror plane perpendicular to the $b$-axis. This symmetry is broken below the structural transition at $T_{\text{S}}\approx79$~K associated with [ND$_4$]$^+$ groups ordering in a staggered pattern along the $b$-axis. Consequently, all the exchange couplings, except the ones connecting the fourth neighbors, split into two symmetry-inequivalent groups, despite having almost identical bond lengths. Such bond splitting may be amplified by pressure and give measurable effects in the dynamical response of the high-pressure phases. 
Our model does not include the antisymmetric Dzyalloshinskii--Moriya interaction (DMI), although it is allowed by symmetry and the spin-current model does require a DMI to generate electric polarization. The strength of such interactions is expected to be very weak, as the polarization observed in experiment is quite small. Strong spin-orbital couplings are required for it to be significant. Fe$^{3+}$ ion has quenched orbital angular momentum in the electronic ground state determined by Hund’s rules, which should lead to extremely weak DMI. Furthermore, a weak DMI would generate cycloids with very long wavelengths (small ordering vectors), but the cycloid observed in {\ruby} has a short wavelength corresponding to $k_c\approx0.23$~r.l.u.. Clearly, DMI cannot be the origin of the cycloidal spin structure in zero field. Instead, the frustrated $J_2$-$J_4$ exchanges are the key to understand the zero-field cycloidal phase as shown in this work. 
A third subtle aspect is the anisotropy within the easy-plane. As mentioned in the beginning, the spin-flop transitions take place at different critical fields when external magnetic fields are applied along $\textbf{\textit{a}}$ and $\textbf{\textit{c}}$. This can be understood by replacing the easy-plane anisotropy with two easy-axis anisotropy in the $ac$-plane. It is a subleading anisotropy that plays an important role in creating distorted cycloids and odd harmonics of magnetic Bragg peaks in the intermediate field regime \citep{Tian-PRB}. We observe a small spin gap ($\sim 0.038(53)$~meV) associated with this weak anisotropy [Supplementary Section S1]. However, its value can not be obtained reliably from our data due to the strong temperature damping effect.
In this work, we conducted a systematic inelastic neutron scattering study of molecule-based multiferroic compound {\ruby} in the zero-field cycloidal phase and the high-field canted antiferromagnetic phase. Even though, the electric polarization measurements of {\ruby} suggests a switch of multiferroic mechanism from the spin current model to the $p$-$d$ hybridization model concurrently with the field-induced transition, our LSWT analysis for both phases arrived at essentially the same magnetic model, implying that magnetic exchange interactions in {\ruby} are much more robust than electric polarization in response to external magnetic fields. 
}

\section*{ACKNOWLEDGMENTS}
We thank Feng Ye and Clarina dela Cruz for valuable discussion, and Andrei Savici for the help in data reduction. X.B. and H.B.C acknowledges the support of US DOE BES Early Career Award KC0402010 under Contract DE-AC05-00OR22725. R.S.F. is supported by the U.S. Department of Energy, Office of Basic Energy Sciences, Materials Sciences and Engineering Division. This research used resources at the High Flux Isotope Reactor and Spallation Neutron Source, a DOE Office of Science User Facility  operated  by  the  Oak  Ridge  National  Laboratory. 
\bibliographystyle{apsrev4-2}
\bibliography{ruby}
%
\clearpage
\onecolumngrid
\setcounter{figure}{0}
\setcounter{table}{0}
\setcounter{equation}{0}
\renewcommand{\thefigure}{S\arabic{figure}}
\renewcommand{\thetable}{S\arabic{table}}
\renewcommand{\theequation}{S\arabic{equation}}
\renewcommand{\thesection}{S\arabic{section}}   

\begin{center}
{\large \bf Supplementary Information}
\end{center}

\section*{S1. Complimentary triple-axis neutron scattering data}
\label{S1}

The CTAX experiment was performed with a fixed final neutron energy of $E_{f} = 3.5$~meV, yielding an elastic energy resolution (HWFM) of $\approx 0.1$~meV. The higher harmonic neutrons were removed by the cooled Be filter placed between the sample and the analyzer. The collimation of guide-open-sample-80'-open was used throughout the experiment.

\begin{figure*} [h]
\centering\includegraphics[width=5in]{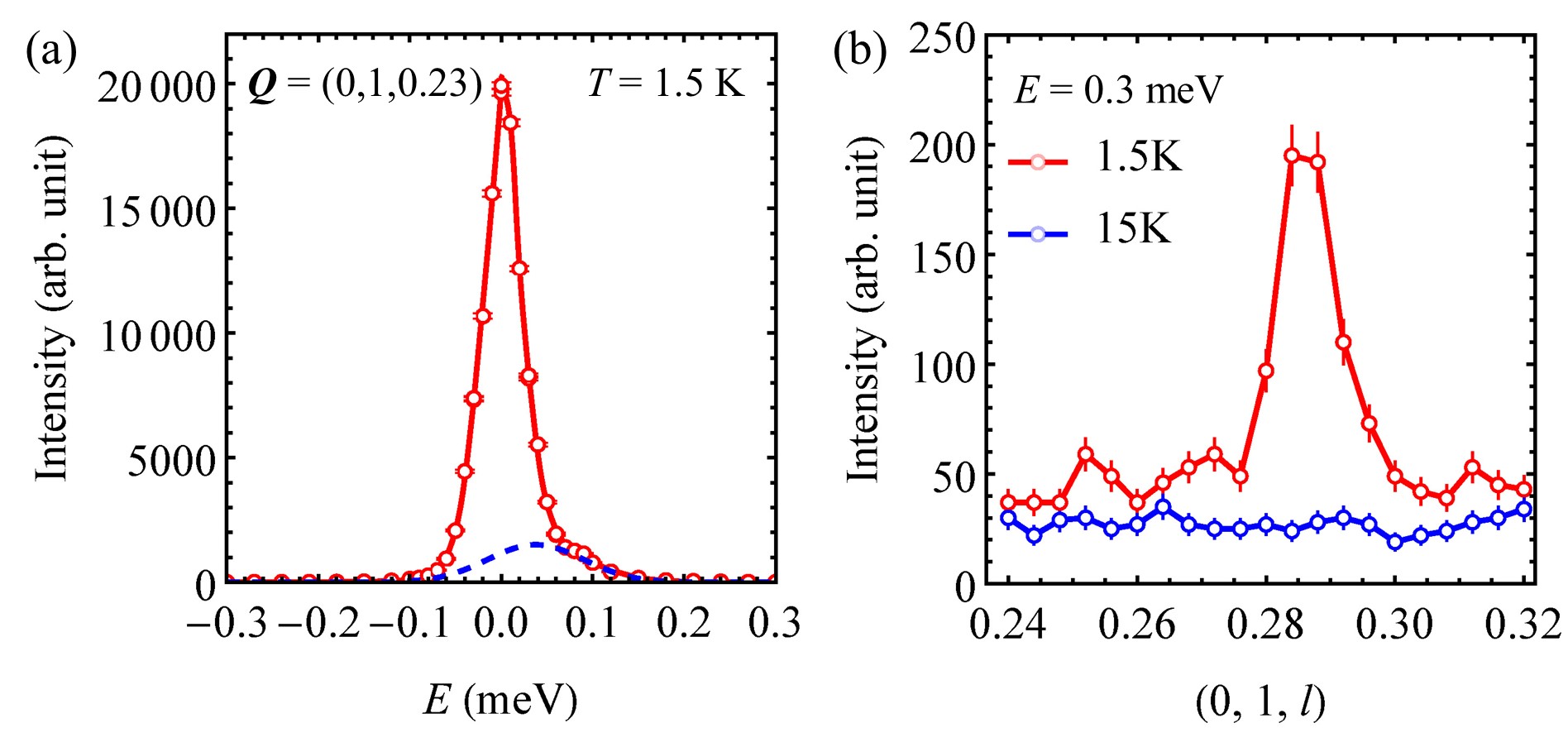}
\caption{\label{fig:CTAX} (a) Constant-$\bf Q$ scan at the magnetic Bragg peak $(0,1,0.23)$. A broad shoulder indicates possible existence of a gap due to anisotropy within the $ac$-plane. The fitted peak (blue dashed curve) is centered around $0.0376$~meV with a FWHM of $0.1237$~meV, much larger than the FWHM ($0.0563$~meV) of the elastic peak (red curve). The true value of the anisotropy gap is difficult to estimate due to the strong temperature damping effect. (b) Temperature dependence of constant-energy scans at $E = 0.3$~meV measured along $(0,1,l)$ at $T = 1.5$ and $15$~K. The lines are guide for the eye.}
\end{figure*} 

\clearpage
\section*{S2. Data processing}

\begin{figure*}[h]
\centering\includegraphics[width=0.8\textwidth]{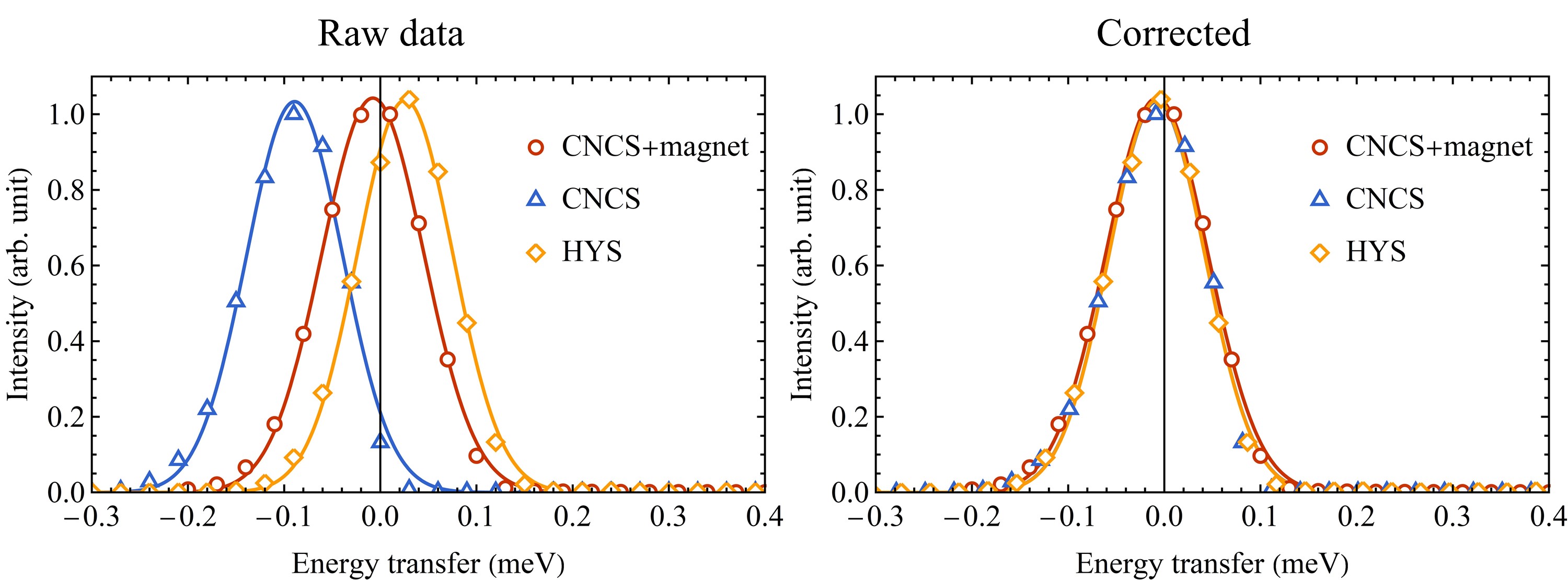}
\caption{\label{fig:elastic} Elastic cuts of three datasets collected in this study at ${\bf Q}=(0,0.5,0)$. The curves are fittings of Gaussian profiles. Note the raw data from the zero-field CNCS experiment and the HYS experiment have non-zero shifts, resulting from imperfect calibration of the time-of-flight in the experiments. It can be corrected in Mantid by adding a time offset. The linear spin-wave analysis is performed on corrected data.}
\end{figure*}

\begin{figure*}[h]
\centering\includegraphics[width=1\textwidth]{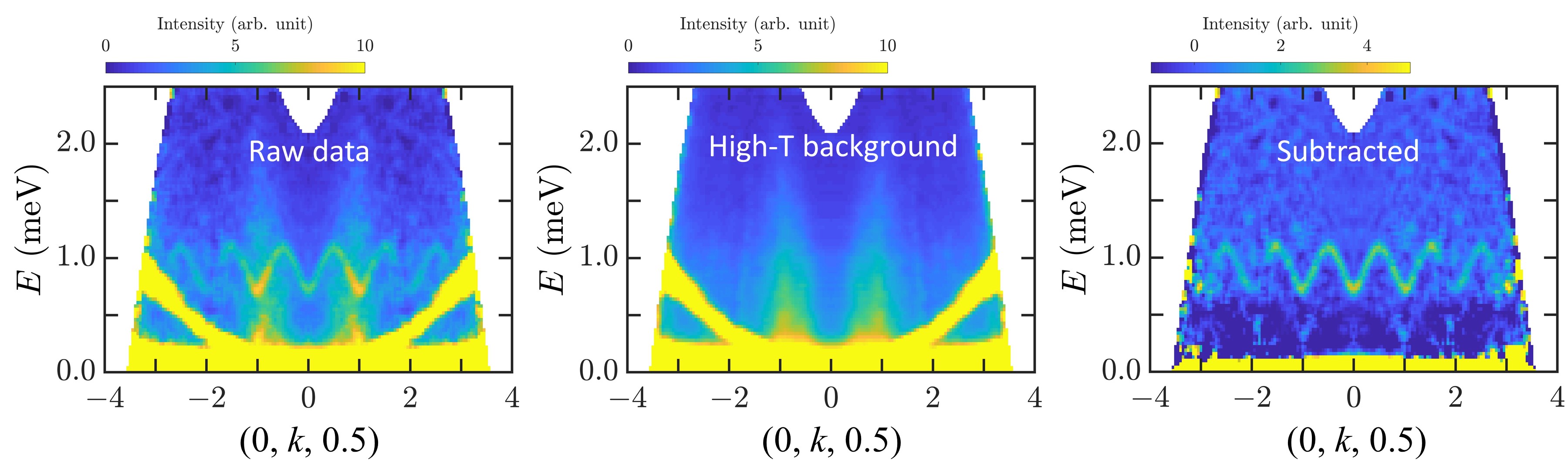}
\caption{\label{fig:bk_sub} The energy-momentum cut at $h=0$~r.l.u. and $l=0.5$~r.l.u., showing the raw data (Left), the high-T ($20$~K) background (Middle) and the background-subtracted data (Right) of the CNCS $6$~T experiment. At the time of the experiment, the CNCS instrument was not yet equipped with a radial collimator. Scattering from the sample environment, primarily the magnet, gave rise to very intensive parabolic-shaped signals and rod-like signals shown in the figure. By subtracting a measurement at $20$~K, magnetic signals can be resolved quite well for almost all slices, but a small degree of over-subtraction below $0.6$~meV can be seen in this cut. Note that the intensity scale is extended to negative numbers to highlight the over-subtracted region. There are clearly some hints of the low-energy branch between $0.3$ and $0.6$~meV.}
\end{figure*}

\clearpage
\section*{S3. Linear spin-wave fitting}

\begin{table}[th]
\begin{tabular}{|r|r|r|r|r|r|r|r|}
\hline
              & $J_1$  & $J_2$  & $J_3$  & $D$    & scale  & $a_\text{bk}$ & $b_\text{bk}$ \\ \hline
$J_1$         & 1.000  & -0.881 & -0.450 & -0.191 & -0.169 & 0.042         & -0.034        \\ \hline
$J_2$         & -0.881 & 1.000  & 0.686  & -0.044 & 0.146  & -0.028        & 0.016         \\ \hline
$J_3$         & -0.450 & 0.686  & 1.000  & 0.084  & 0.107  & -0.040        & 0.026         \\ \hline
$D$           & -0.191 & -0.044 & 0.084  & 1.000  & 0.098  & -0.025        & 0.018         \\ \hline
scale         & -0.169 & 0.146  & 0.107  & 0.098  & 1.000  & -0.293        & 0.200         \\ \hline
$a_\text{bk}$ & 0.042  & -0.028 & -0.040 & -0.025 & -0.293 & 1.000         & -0.936        \\ \hline
$b_\text{bk}$ & -0.034 & 0.016  & 0.026  & 0.018  & 0.200  & -0.936        & 1.000         \\ \hline
\end{tabular}
\caption{Correlation matrix for the zero-field model parameters.}
\label{tab:zero-field}
\end{table}

\begin{table}[th]
\begin{tabular}{|r|r|r|r|r|r|r|r|r|r|}
\hline
              & $J_1$  & $J_2$  & $J_3$  & $J_4$  & $J_5$  & $D$    & scale  & $a_\text{bk}$ & $b_\text{bk}$ \\ \hline
$J_1$         & 1.000  & -0.881 & 0.162  & -0.303 & -0.489 & -0.028 & -0.048 & 0.001         & -0.004        \\ \hline
$J_2$         & -0.881 & 1.000  & -0.346 & 0.296  & 0.509  & 0.064  & 0.013  & 0.012         & -0.008        \\ \hline
$J_3$         & 0.162  & -0.346 & 1.000  & 0.698  & 0.102  & -0.184 & 0.042  & -0.002        & -0.004        \\ \hline
$J_4$         & -0.303 & 0.296  & 0.698  & 1.000  & 0.213  & 0.037  & 0.012  & 0.007         & -0.008        \\ \hline
$J_5$         & -0.489 & 0.509  & 0.102  & 0.213  & 1.000  & 0.027  & 0.050  & 0.007         & -0.009        \\ \hline
$D$           & -0.028 & 0.064  & -0.184 & 0.037  & 0.027  & 1.000  & -0.039 & 0.018         & -0.015        \\ \hline
scale         & -0.048 & 0.013  & 0.042  & 0.012  & 0.050  & -0.039 & 1.000  & -0.238        & 0.129         \\ \hline
$a_\text{bk}$ & 0.001  & 0.012  & -0.002 & 0.007  & 0.007  & 0.018  & -0.238 & 1.000         & -0.919        \\ \hline
$b_\text{bk}$ & -0.004 & -0.008 & -0.004 & -0.008 & -0.009 & -0.015 & 0.129  & -0.919        & 1.000         \\ \hline
\end{tabular}
\caption{Correlation matrix for the high-field model parameters.}
\label{tab:high-field}
\end{table}

\begin{figure*}[h]
\centering\includegraphics[width=0.8\textwidth]{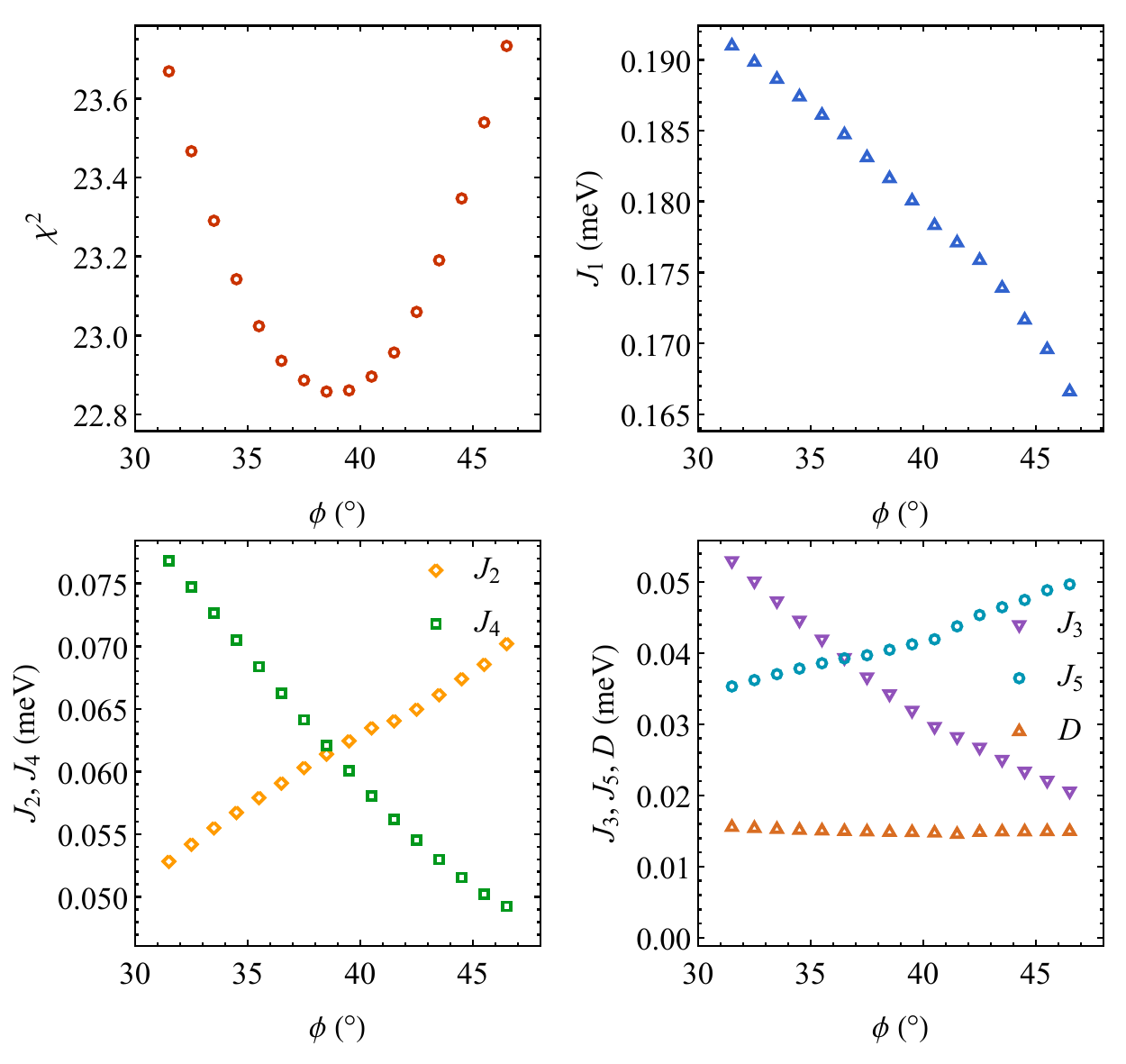}
\caption{\label{fig:SI_scan_phi} The reduced $\chi^2$ and the optimal model parameters as a function of the phase difference $\phi$, obtained from fitting the zero-field data. The minimal $\chi^2$ corresponds to a phase difference $\phi\approx 39^\circ$, reasonably close to the value reported in neutron diffraction studies, $41.5^\circ$. }
\end{figure*}

\clearpage
\section*{S4. Comparison between data and LSWT modeling}

To elucidate the overall quality of the modeling, we show detailed cuts across reciprocal space comparing the zero-field data (Fig. S5-S7) and $B=6$~T data (Fig. S8-S10) with respective best LSWT fits. See main text for details of instrumental setup. The fitted background functions are not plotted for clarity.

\begin{figure*}[h]
\centering\includegraphics[width=1\textwidth]{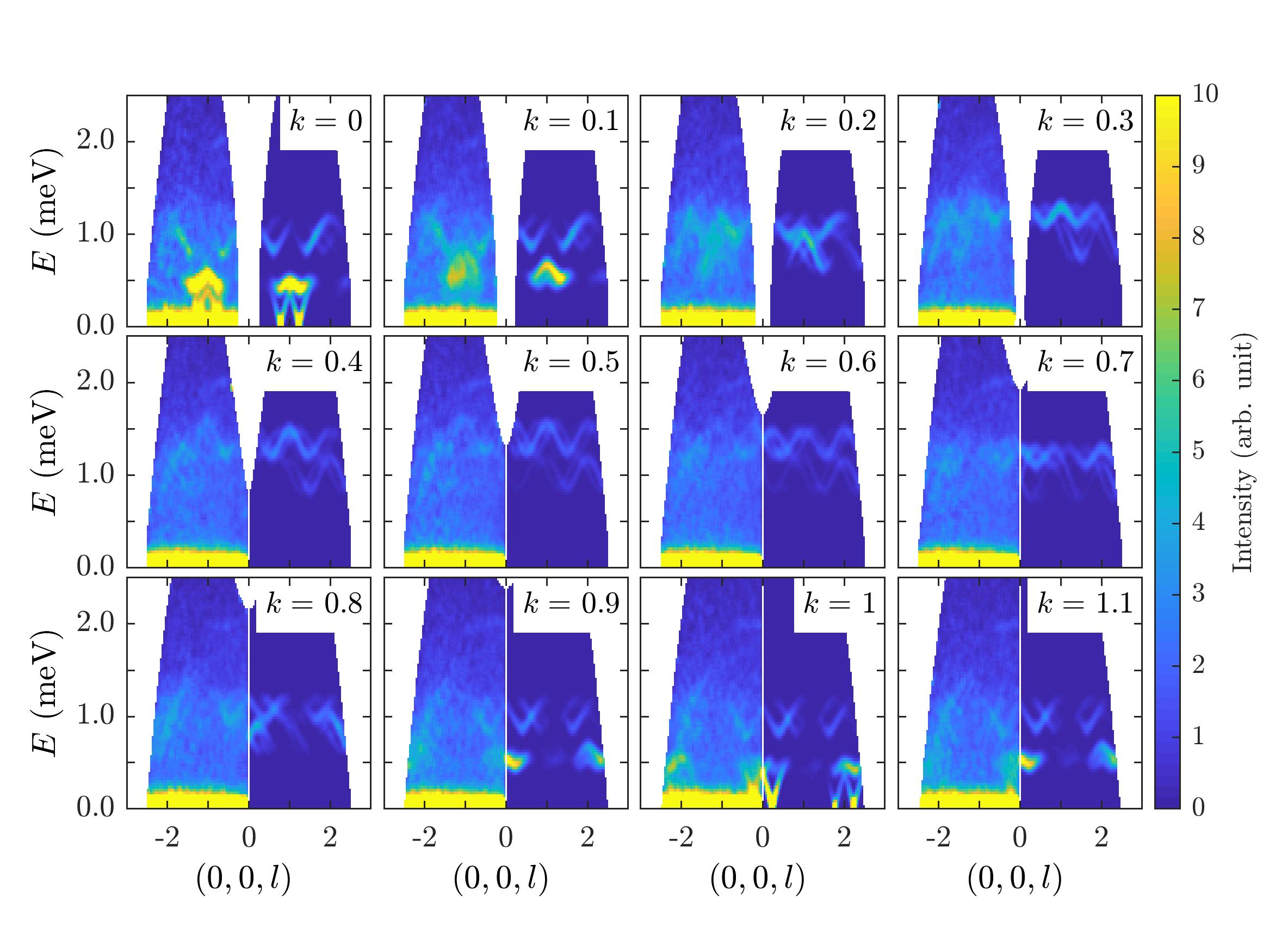}
\caption{Comparison between the zero-field data in the $(0,k,l)$-plane and its best fitted model. In each panel, the data is shown on the left and the fit on the right, integrated over $|\Delta k|\le 0.05$~r.l.u. and $|h|\le 0.05$~r.l.u.. }
\end{figure*}

\begin{figure*}[h]
\centering\includegraphics[width=1\textwidth]{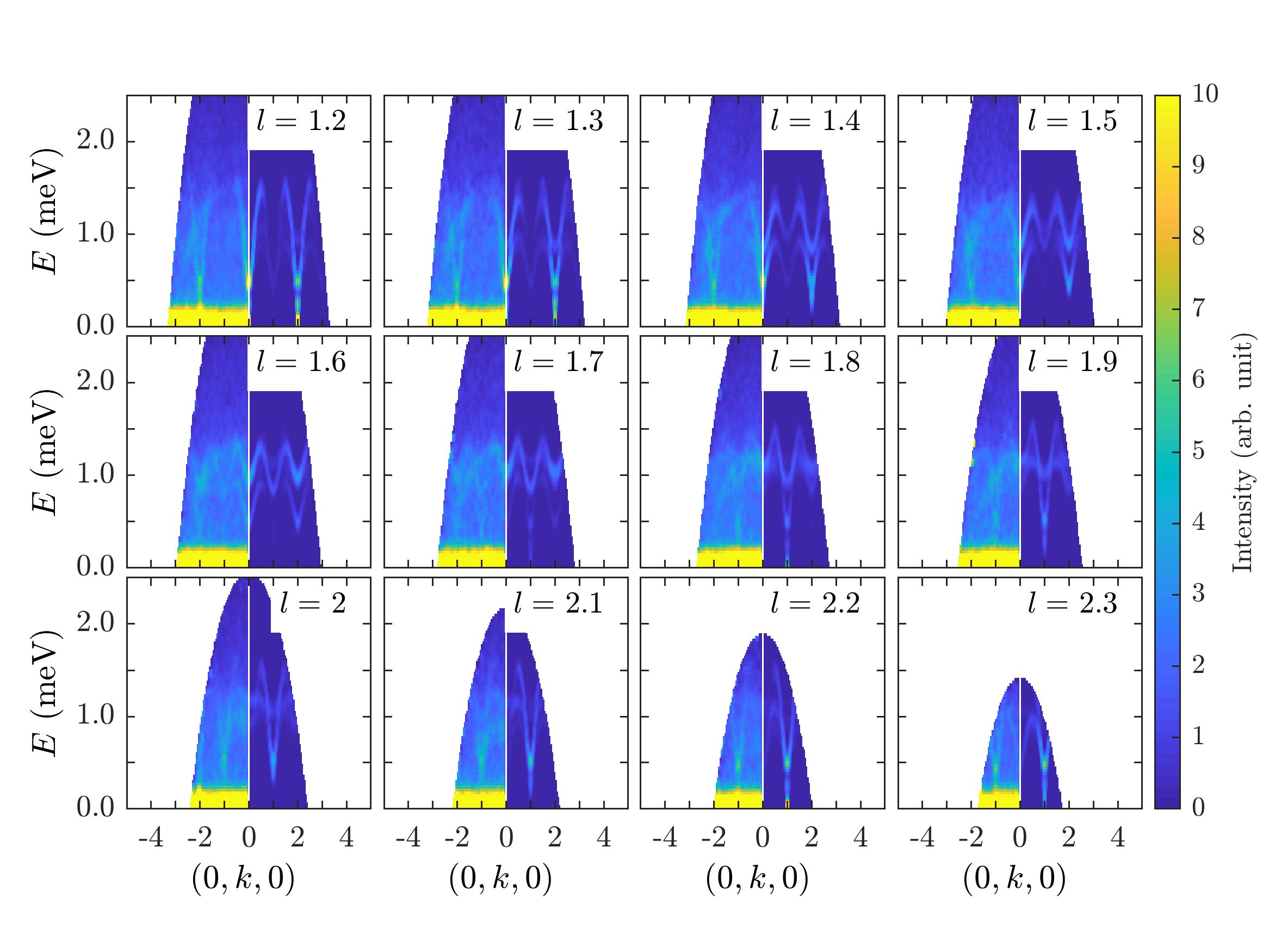}
\caption{Comparison between the zero-field data in the $(0,k,l)$-plane and its best fitted model. In each panel, the data is shown on the left and the fit on the right, integrated over $|\Delta l|\le 0.05$~r.l.u. and $|h|\le 0.05$~r.l.u.. }
\end{figure*}

\begin{figure*}[h]
\centering\includegraphics[width=1\textwidth]{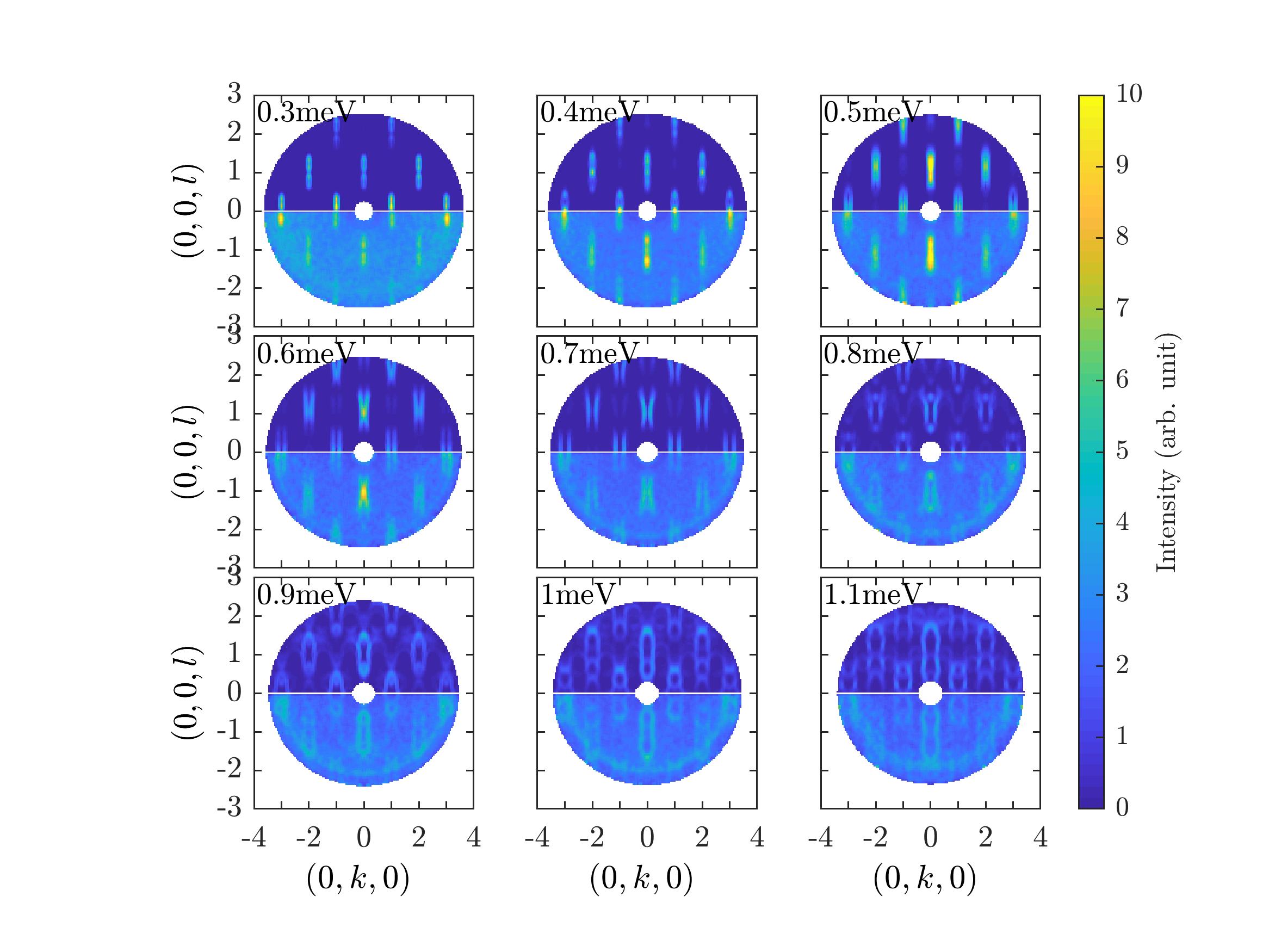}
\caption{Momentum-dependence of excitation spectra of the zero-field data and the model integrated over $|\Delta E|\le 0.05$~meV at selected energies indicated at the top-left corner of each panel. The data is shown on the bottom half of each panel and the fit on the top half.}
\end{figure*}

\begin{figure*}[h]
\centering\includegraphics[width=1\textwidth]{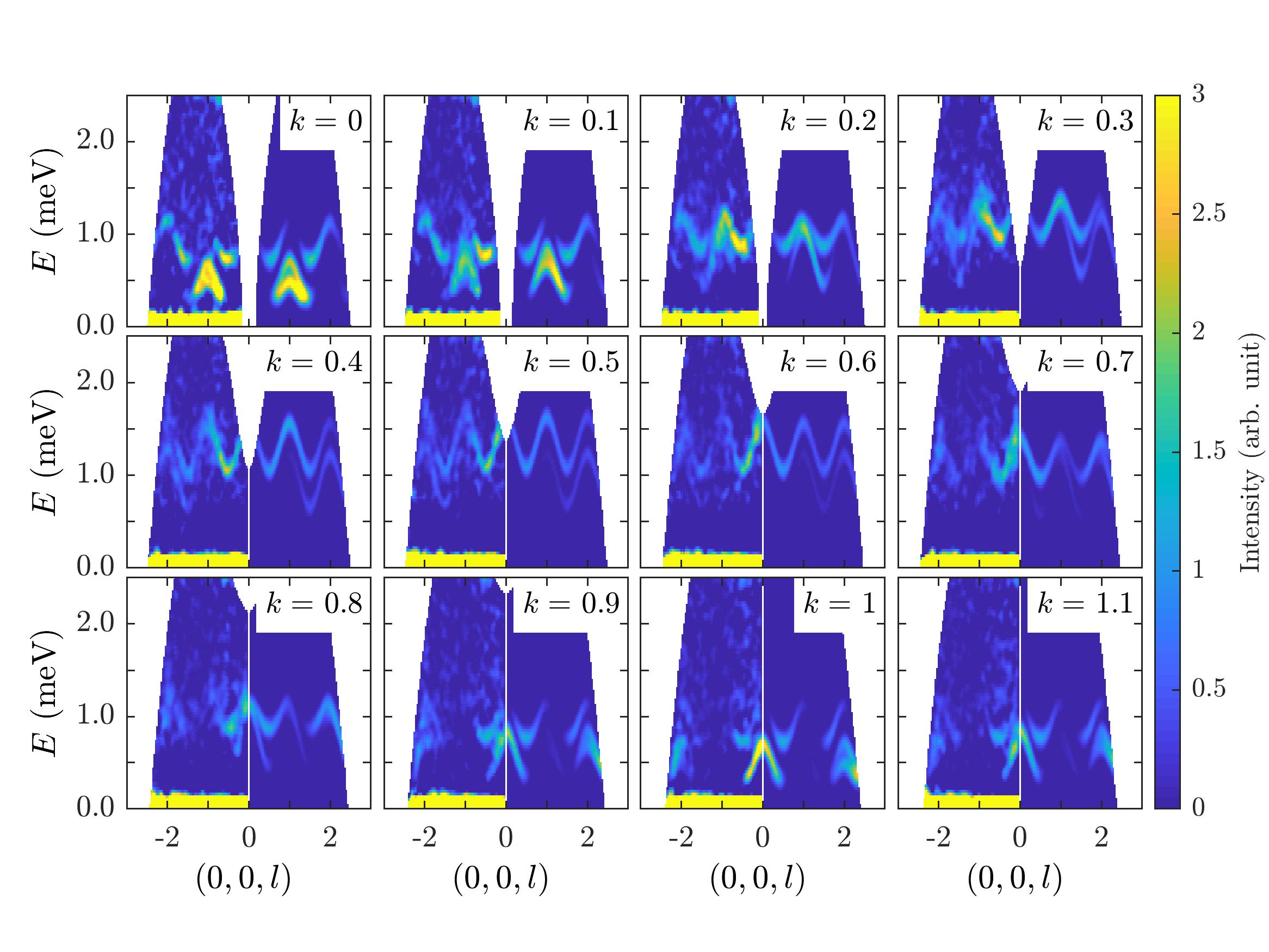}
\caption{Comparison between the $B=6$~T data in the $(0,k,l)$-plane and its best fitted model. In each panel, the data is shown on the left and the fit on the right, integrated over $|\Delta k|\le 0.05$~r.l.u. and $|h|\le 0.05$~r.l.u.. }
\end{figure*}
\begin{figure*}[h]
\centering\includegraphics[width=1\textwidth]{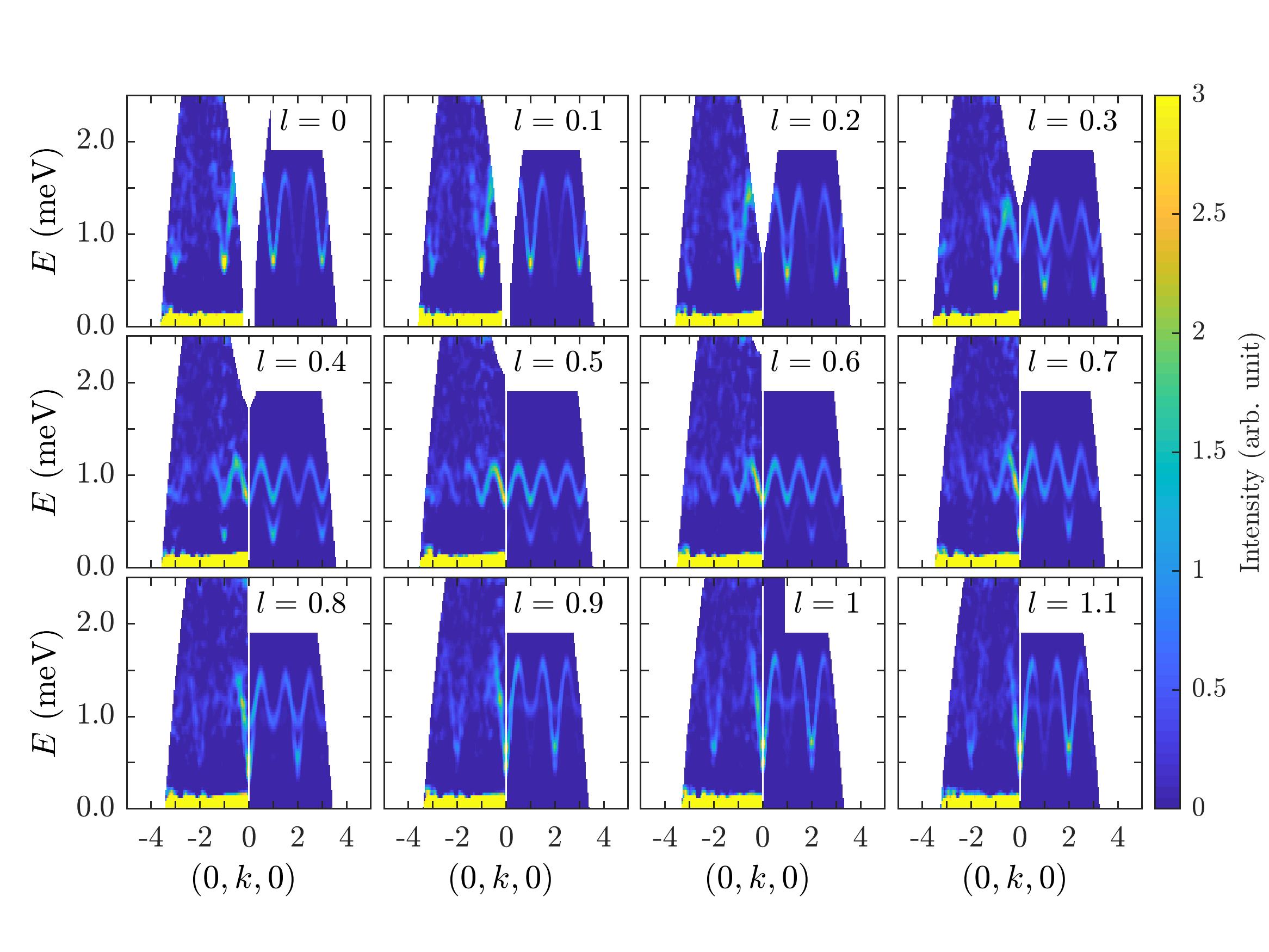}
\caption{Comparison between the $B=6$~T data in the $(0,k,l)$-plane and its best fitted model. In each panel, the data is shown on the left and the fit on the right, integrated over $|\Delta l|\le 0.05$~r.l.u. and $|h|\le 0.05$~r.l.u.. }
\end{figure*}

\begin{figure*}[h]
\centering\includegraphics[width=1\textwidth]{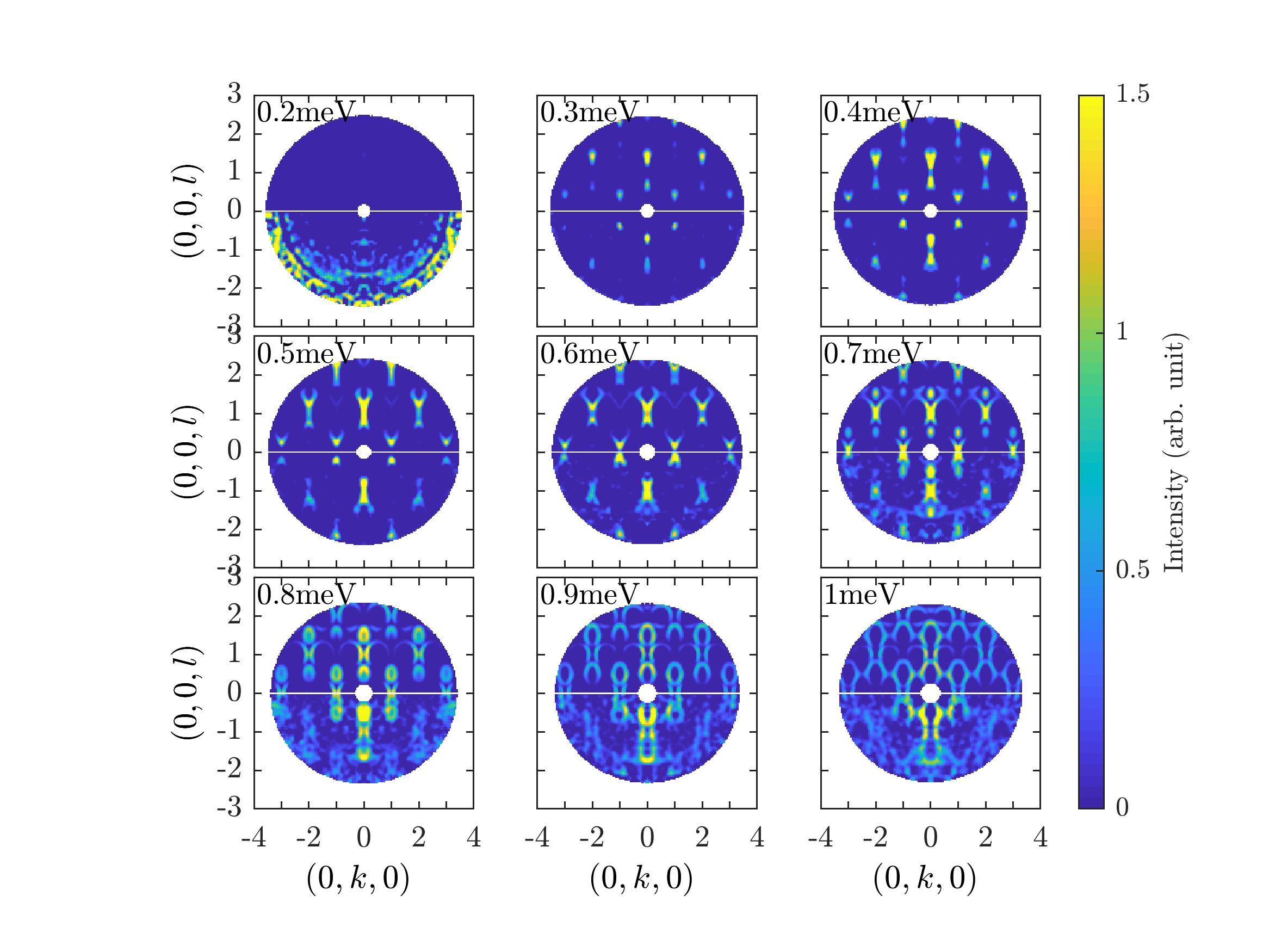}
\caption{Momentum-dependence of excitation spectra of the $B=6$~T data and the model integrated over $|\Delta E|\pm0.05$~meV at selected energies indicated at the top-left corner of each panel. The data is shown on the bottom half of each panel and the fit on the top half.}
\end{figure*}

\end{document}